\documentstyle[epsf,amssymb,12pt]{article}
\newcommand{\beq}{\begin{equation}}
\newcommand{\eeq}{\end{equation}}
\newcommand{\bqa}{\begin{eqnarray}}
\newcommand{\eqa}{\end{eqnarray}}
\newcommand{\bm}[1]{\mbox{\bf #1}}
\newcommand{\x}{x}

\newcommand{\eg}{{\frenchspacing\em e.\hspace{0.4mm}g.{}}}
\newcommand{\ie}{{\frenchspacing\em i.\hspace{0.4mm}e.{}}}
\newcommand{\eq}[1]{{\frenchspacing Eq.~\ref{#1}}}
\title{{\small \hfill WUB  95-21}\\
       {\small \hfill HLRZ 45/95}\\[1cm]
Hyper-Systolic Processing on APE100/Quadrics:\\
$N^2$-Loop Computations}
\author{Th. Lippert$^a$, G. Ritzenh\"ofer$^a$,
        U. Gl\"assner$^b$, H. Hoeber$^a$,
        A. Seyfried$^b$, and K. Schilling$^{a,b}$\\[0.2cm]\footnotesize
$^a$HLRZ, c/o KFA-J\"ulich, D-52425 J\"ulich,
    Germany and DESY, Hamburg, Germany\\[-0.2cm] \footnotesize
$^b$Department of Physics, University of Wuppertal,
    D-42097 Wuppertal, Germany}

\date{}

\begin{document}
\maketitle

\abstract{We investigate the performance gains from hyper-systolic
  implementations of $n^2$-loop problems on the massively parallel
  computer Quadrics, exploiting its 3-dimensional interprocessor
  connectivity.  For illustration we study the communication aspects
  of an exact molecular dynamics simulation of $n$ particles with
  Coulomb (or gravitational) interactions.  We compare the
  interprocessor communication costs of the standard-systolic and the
  hyper-systolic approaches for various granularities.  We predict
  gain factors as large as 3 on the Q4 and 8 on the QH4 and measure
  actual performances on these machine configurations.  We conclude
  that it appears feasable to investigate the thermodynamics of a full
  gravitating n-body problem with $O(10000)$ particles using the new
  method on a QH4 system.}

\parindent 0pt
\parskip 12 pt
\section{Introduction}

Many grand challenge tasks in computational science involve the
solution of $n^2$-loop problems, demanding for algorithms and
techniques viable on parallel machines. One would wish to increase the
number of processors proportional to $n^{\alpha}$ such as to achieve a
computational complexity of $O(n^{2-\alpha})$ per
processor\footnote{In Ref.~\cite{PARISI}, Parisi discusses the
  limiting case of $O(const.)$ complexity per processor on computers
  with arbitrarily large parallelism.}.  This strategy sets the goal
towards the limit of {\it low} granularity!

In this paper we treat, as a prototype of an $n^2$-loop
problem\footnote{Closely related problems are protein folding, polymer
  dynamics with long range forces, signal processing, or fully coupled
  neural nets.}, the molecular dynamics of $n$ equal particles with
long range Coulomb forces. Computationally, this implies repeated
calculation of all interparticle distances $r_{ij}$ within the full
system.  The state-of-the-art system size in molecular dynamics with
an exact treatment of such long-range forces is at present of
$O(10.000)$, requiring machines in the 10 GFlops range.

On a massively parallel computer with $p$ processors and distributed
memory, the coordinates of the particles are assigned to different
processors and therefore interprocessor movement of data becomes a
considerable cost factor in the computation of two particle forces.
In the limit of low granularity, \ie\ $p \simeq n$, interprocessor
communication is in fact the major performance bottleneck. This holds
particularly true for machines with next-neighbour interprocessor
communication networks as \eg\ the APE100/Quadrics system,
manufactured by Alenia Spazio S.P.A.\ \cite{ALENIA} or the 16k node
machine under construction at Columbia University \cite{COLUMBIA}.  It
is therefore crucial to gear algorithms to render low communicational
overhead on such machines for computing $n^2$-loop problems.

An earlier proposal to implement $n^2$-loop systems on the massively
parallel computer Quadrics has been presented by
Paolucci\cite{PAOLUCCI}.  His parallel approach rests upon the
replicated data method\cite{REPLICATED} in conjunction with global
summations over the whole machine and arrives at a total
communicational complexity of $O(np^{\frac{4}{3}})$.

In this paper we will fathom the power of systolic array computation.
The latter is a rather general technique to organize data flow on
parallel systems that implies the potentiality to improve the
interprocessor communication complexity for $n^2$-loop problems.  In
systolic loop processing \cite{PETKOV}, on a $p$ processor machine
working on $n$ particle dynamics with long range forces, the latter is
of order $O(np)$, as each of the $n$ data elements has to be
communicated to $p$ processors.

Recently, we have introduced the concept of hyper-systolic array
pro\-cess\-ing\cite{HYPER}. This novel parallel algorithm is an
extension of the standard-systolic method.  In this approach, the data
flow is reorganized such that the interprocessor communication is
reduced to order $2n(2\sqrt{\frac{p}{2}}-1))$ for an array of $n$
particles, at no additional computational cost.  This is accomplished
at the expense of only a modest amount of additional storage.

Hyper-systolic data piping is structurally related to the so-called
postage stamp problem in Additive Number Theory which can serve to
classify all possible hyper-systolic flow patterns.  Needless to say
the hyper-systolic concept is very general and can be applied to all
sorts of $n^2$-computations, as \eg\ matrix
computations\cite{GALLI,MATRIX}.

In this work we shall benchmark the hyper-systolic concept for solving
$n^2$-loop problems on the Quadrics system.  For assessing performance
gains, we choose the 1-dimensional standard-systolic method as our
reference point. We will present an efficient mapping of 1-dimensional
systolic arrays onto the 3-dimensional network of the Quadrics.  In
our setup, one hyper-systolic movement is performed in $\le 3$
interprocessor communication steps.

\section{Systolic Array Implementation\label{SYS}}
We consider for definiteness the computation of the local Coulombic forces
within an n-body problem
\beq
\vec y_i = \vec
F(\vec x_i) = \sum_{{j=1}\atop{j\ne i}
}^n \frac{\vec x_i - \vec x_j} {|\vec x_i -
\vec x_j|^3}, \qquad i=1,\dots,n,
\label{NEWTON}
\eeq
the evaluation of which represents the computational bottleneck.

\begin{figure}[htb]
\centerline{\epsfxsize=.8\textwidth\epsfbox{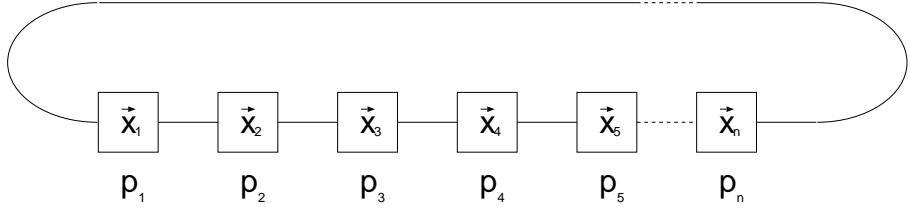}}
\caption[Systolic array.]{
  Systolic array GRA.
\label{SYSTOLIC}}
\end{figure}

Let us assume for a start the limit of granularity $g$ to be $g = n/p
= 1$.  Our standard-systolic loop approach to compute \eq{NEWTON}
proceeds by mapping the processors onto a 1-dimensional periodic chain
(GRA\footnote{This name was chosen as a mnemonic to the 20th century
  Roman traffic solution {\it Grande Raccordo Anulare}.}), see
Fig.~\ref{SYSTOLIC}.
The information on the location of the physical system in
configuration space, ${\bm{x}}=(\vec x_1, \vec x_2,\dots,\vec x_n)$,
is distributed over the compute system such that each processor $p_i$
holds in its local memory the coordinate $\vec x_i$ as its `resident
vector'.  Besides the local resident vector $\vec x_i$ there is a
local migrating vector $\vec{\x}_i'$, that originally coincides with
$\vec x_i$.  During a systolic step, $\vec{\x}_i'$ is shifted
cyclically onto the right hand next-neighbour processor on the
1-dimensional systolic chain GRA. After completion of this systolic
move, the computation of the local pairings is performed in parallel
and the results are accumulated into a resident result array
${\bm{y}'}$.  After $n-1$ such systolic steps, \eq{NEWTON} is fully
computed\footnote{One invokes Newton's {\it actio = reactio} to reduce
  the amount of computation by a factor two.}.

\begin{figure}[p]
\centerline{\epsfxsize=.8\textwidth\epsfbox{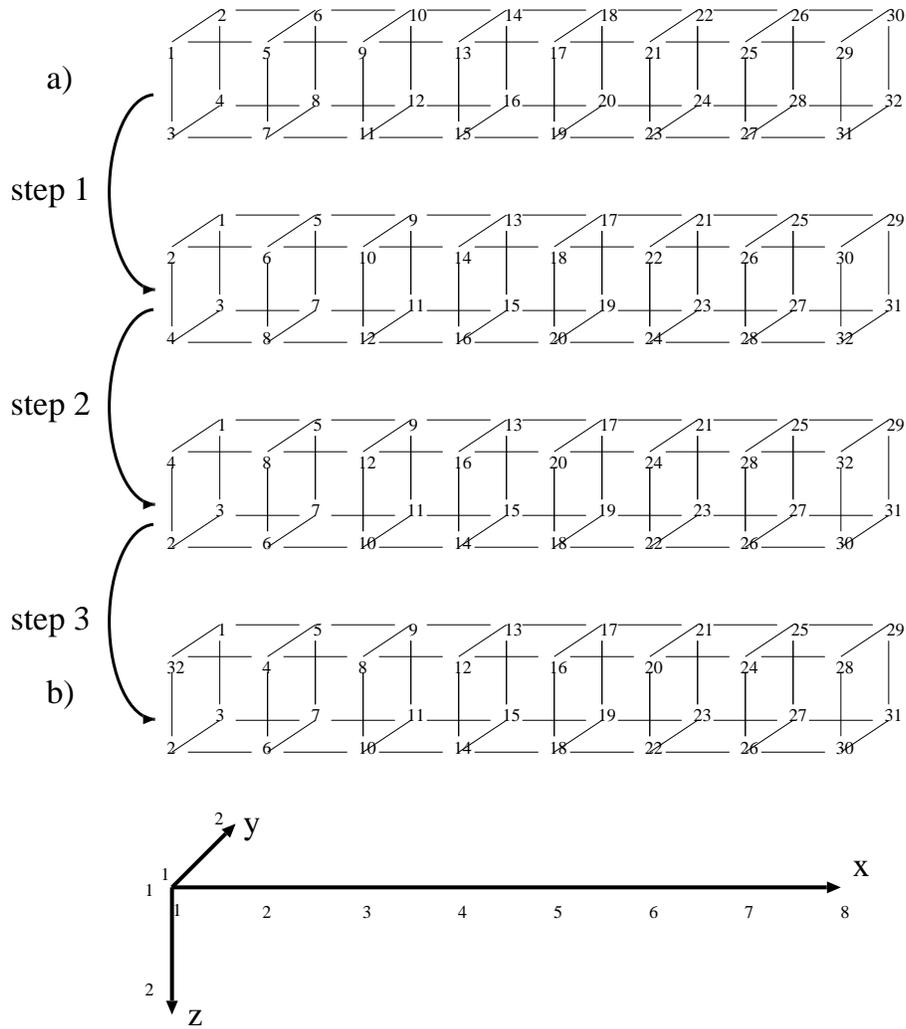}}
\caption[]{
  Systolic movement.  a) Initial state: Mapping of a 1-dimensional
  systolic array with $n=32$ onto the 3-dimensional Quadrics Q4 with
  $p=32$ processors. The systolic chain is periodically closed.  b)
  Final state: reached after 3 steps.  The coordinate system is given
  to specify the Quadrics' $8\times 2\times 2$ processor grid.
\label{QUAD_TOP}}
\end{figure}

In Fig.~\ref{QUAD_TOP}a, we illustrate the mapping of the
1-dimensional systolic array onto the Quadrics processor topology: on
a Quadrics Q4, with $n=p=32$.  The processors of the Q4 are arranged
in a $8\times 2\times 2$-grid that is toroidally closed\footnote{The
  mapping of systolic arrays with $n=p$ onto other Quadrics
  configurations, like Q16, QH2, and QH4 is straightforward.}.

One must differentiate between a movement on the logical, systolic
chain GRA and its actual hardware realization, in terms of
interprocessor communications: one systolic movement to the next
neighbour processor on GRA amounts to three steps on the
interprocessor links, as depicted in Fig.~\ref{QUAD_TOP}:
\begin{description}
\item[Step 1:] cyclical shift to next neighbour in $y$-direction,
\item[Step 2:] cyclical shift to next neighbour in $z$-direction, at
  $y=1$ only,
\item[Step 3:] cyclical shift to next neighbour in $x$-direction,
  $y=1$, $z=1$ only.
\end{description}
For the entire systolic process, $3(p-1)$ such interprocessor communication
operations are needed. Therefore the communicational complexity of
standard-systolic computations on the Quadrics is $3p(p-1)$ .

Proceeding to the case of $n>p$, we partition the systolic array into
$\frac{n}{p}$ subarrays and distribute each subarray across the
machine. From the point of view of interprocessor communication the
pairings of coordinates from different subarrays present no additional
complication.  The total number of interprocessor communication
operations is now found as $3n(p-1)$, while the total number of
computation operations remains unaltered, $n(n-1)$.

\section{Hyper-Systolic Algorithm\label{HYP}}
The hyper-systolic reorganization of the computation of \eq{NEWTON}
will enable us to reduce the complexity of interprocessor
communication to $O(np^\frac{1}{2})$; thus, interprocessor
communication ultimately becomes negligible compared to the pure
computational complexity, on machines with a very large number of
processors.

Quadrics machines allow for an efficient realization of hyper-systolic
array computing, due to their 3-dimensional cubic
connectivity\footnote{From the point of view of 1-dimensional systolic
  data arrangement and movement the 3-dimensional network of the
  Quadrics supports accelerated communication along transversal
  directions (wormhole communication).}.

\subsection{Definition: Regular Base Algorithm}

The hyper-systolic data movement generalizes the standard-systolic one
by increasing the number of resident arrays from one to $k$, with
$k\ll p$ and introducing cyclical shifts with strides $a_t \ge 1$.

Let us  give an operational description of the hyper-systolic
algorithm for extreme granularity  $g = 1$:
\begin{enumerate}
\item For a given array $\bm{x}$ of length $n$, $k$ arrays
  $\hat{\bm{x}}_{t}$ are generated by shifting the original array
  $\bm{x}$ $k$ times by strides $a_t$, and copying $\bm{x}$ to
  $\hat{\bm{x}}_{t}$ in step $t$, $1\le t\le k$.  The variable stride
  $a_t$ of the shifts must be chosen such that
\begin{enumerate}
\item
all pairings of data elements occur at least once, {\it and}
\item
the number of equal pairings  as well as
\item
the number of shifts $k$ as well as
\item
the number of different strides $a_t$
\end{enumerate}
are minimized.
\item The required $n(n-1)/2$ results $\vec y_i$ in \eq{NEWTON} are
  successively computed and are added to $k$ resulting arrays
  $\hat{\bm{y}}_{t}$.
\item Finally, applying the inverse shift sequence, the arrays
  $\hat{\bm{y}}_{t}$ are shifted back to their proper locations and
  corresponding entries are summed up to build the elements of the
  final result array $\bm{y}$.
\end{enumerate}

The interested reader can find the optimal solutions to this problem
for $8\leq p \leq 64$ in Ref.~\cite{HYPER}. In general, one can resort
to so-called regular base solutions which are very simple yet still in
the optimal complexity class.

In the following we implement the hyper-systolic algorithm using the
regular base:
\beq
\arraycolsep=0pt
\begin{array}{cccccccc}
A_{k=2K-1} &=& \Big( 0,& 1,1,\dots,1                 &,&\, K,K,\dots,K
        & \Big) &,\\
           & &       & \raisebox{.3cm}{$\underbrace{\makebox[1.6cm]{}}$}&
                     & \raisebox{.3cm}{$\underbrace{\makebox[1.6cm]{}}$} &
 &  \\
           & &       & K-1                           & & K \\
\end{array}
\label{REGULAR}
\eeq
with $K-1$ shifts by stride $a_t = 1$ and $K$ shifts by stride  $a_t = K$.  The
length of the
base is $k=2K+1$. The shift constant $K$ is given by
$K=\sqrt{\frac{p}{2}}$.

\subsection{Implementation on the Q4.}

\begin{figure}[htb]
  \centerline{\epsfxsize=.8\textwidth\epsfbox{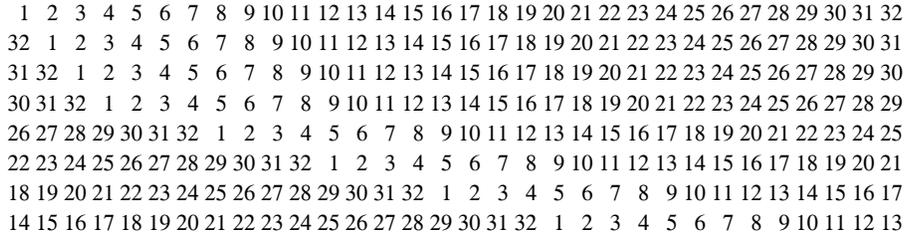}}
\caption[dummy]{Hyper-systolic data movement on the Q4 for $n=32$ using the
regular base
  $A_7=(0, 1, 1, 1, 4, 4, 4, 4)$.\label{HYSYS}}
\end{figure}

For illustration we consider the case $n=32$.  The regular
hyper-systolic base is given by \beq A_7 = \Big(0, 1,1,1,4,4,4,4
\Big), \eeq leading to the storage of eight arrays as depicted in
Fig.~\ref{HYSYS}.
By inspection one finds all pairings of the numbers $1$ to $32$ within
the columns, \ie\ within the 32 local memories.

\begin{figure}[htb]
\centerline{\epsfxsize=.8\textwidth\epsfbox{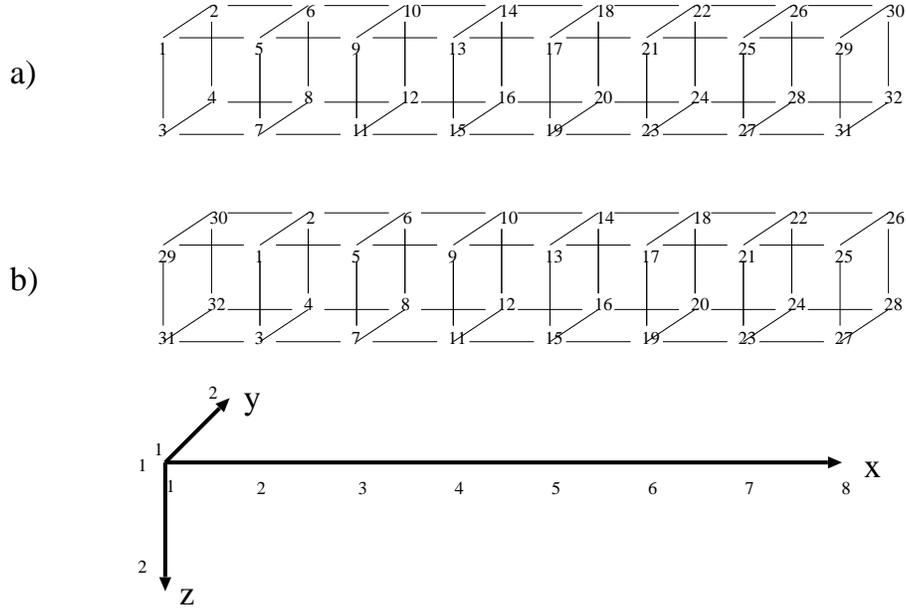}}
\caption[dummy]{Shift by a stride of $4$ on the Quadrics Q4.
a) Initial state,
b) final state.\label{STRIDE}}
\end{figure}

The systolic chain ${\bm{x}}$ is mapped in the very same way as in
section \ref{SYS} for stride $a_t=1$.  The shifts by the stride $K=4$
present a novel feature of the hyper-systolic method: from
Fig.~\ref{STRIDE} it can be seen that such a hyper-shift, on the Q4,
is performed by just {\em one} communication operation to the
next-neighbour in $x$-direction.

Thus the communicational complexity of the regular base hyper-systolic
method on the Q4 is improved compared to the scaling law in
Ref.~\cite{HYPER}, that has been derived for constant communicational
cost for all strides, and is given by $2n(4\sqrt{\frac{p}{2}}-1)$.
The improvement with respect to the standard-systolic method on the Q4
is \beq R=\frac{3n(p-1)}{2n(4\sqrt{\frac{p}{2}}-1)} \simeq
3\sqrt{\frac{p}{32}} =3.  \eeq This result holds also for the case
$n>p$ where we partition the array into $\frac{n}{p}$ subarrays and
compute all relevant interactions\footnote{Recently, the so called
  Half-Orrery algorithm has been presented in Ref.~\cite{SCHOLL}. This
  method also belongs to the class of hyper-systolic algorithms. In
  principle, it can, as is the case with the approach presented here,
  make usage of the symmetry inherent in \eq{NEWTON}. The
  communication, however, is not reduced compared to the
  standard-systolic way, and moreover on the Quadrics, additional
  back-shifts have to be performed. Thus Half-Orrery is not
  competitive on Quadrics.}.

\subsection{Implementation on the QH2 and QH4\label{IMPQH4}}

The 256-node Quadrics QH2 is arranged as an $8 \times 4\times 8$
lattice. The procedure underlying \eq{REGULAR} needs to be generalized
as $\sqrt{128}$ is not integer.  We can utilize the decomposition
$\frac{p}{2}=KK'$ and select the base
\beq \arraycolsep=0pt
\begin{array}{cccccccc}
A_{k=K-1+K'} &=& \Big(0, & 1,1,\dots,1                 &,&\, K,K,\dots,K
          & \Big) &,\\
           & &       & \raisebox{.3cm}{$\underbrace{\makebox[1.6cm]{}}$}&
                     & \raisebox{.3cm}{$\underbrace{\makebox[1.6cm]{}}$} &
 &  \\
           & &       & K-1                           & & K' \\
\end{array}.
\label{REGULAR2}
\eeq
For the QH2, the favourite base reads \beq A_{23} =
\Big(0,1,1,1,1,1,1,1,8,8,8,8,8,8,8,8,8,8,8,8,8,8,8,8 \Big).  \eeq Note
that a stride by 8 can be performed with 2 communication operations
only, analogous to the shifts on the Q4. It turns out that the ratio R
between standard-systolic and hyper-systolic data communication
expense theoretically is about 7.2.

The QH4 with 512 processors and $8\times 8 \times 8$ topology readily
lends itself to the regular hyper-systolic base, \eq{REGULAR}, again
with a stride of $K=16$. Constrained by the topology, a shift with a
stride of $16$ has to be accomplished within 3 communication
operations by 2 strides of length 8, and theoretically a gain in
communicational expense of a factor of $8$ can be achieved.  It is
interesting to note that asymmetric processor grids can increase the
hyper-systolic performance: \eg, if the $16\times 8\times 4$-topology
is realized, the gain factor is about $10$. With a $4\times 4\times
32$-topology, the gain is optimized further to 12.

\section{Performance Results}

\subsection{Performance Tests on the Q4\label{PERFORMANCE}}

The hyper-systolic computation of the interaction given by \eq{NEWTON}
on Qadrics machines has been developed and first tested on the 32-node
Q4 at Wuppertal University, Germany.  In order to speed up the purely
computational part of the program we have employed block matrices in
form of register declarations for the compute intensive part. As the
number of registers is limited to 128 per processor, this approach
amounts to an additional second blocking in form of loop-unrolling in
the number of particles per processor $\frac{n}{p}$. Thus, the array
of $\frac{n}{p}$ elements is subpartitioned into $q$ parts.  In our
three-dimensional molecular dynamics example with Coulombic forces we
are able to perform computations on $2$ arrays of up to $6$ particles,
\ie, $6 \times 6$  coordinate pairings  can be treated  while
no recourse to the local memory is needed. We emphasize that both the
standard-systolic and the hyper-systolic methods use the identical
optimized code concerning this local computational part.

We have measured the total run times, $t_{total}=t_{cpu}+t_{ipc}$, and
the pure interprocessor communication times $t_{ipc}$ as a function of
the total number of particles $n$.

\begin{figure}[htb]
\centerline{\epsfxsize=8cm\epsfbox{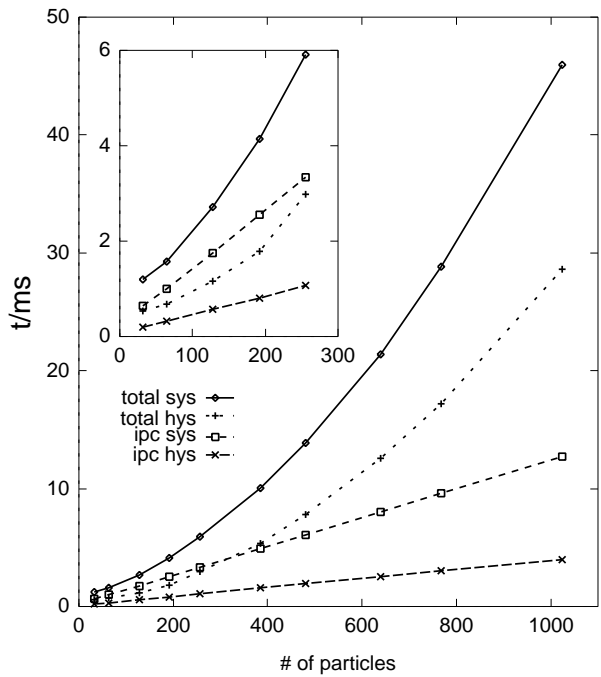}}
\caption[Times]{Overall time (total) and  communication time (ipc) vs.\
  the number of particles for one interaction time-step on the
  32-nodes Quadrics Q4 for the systolic and the hyper-systolic method.
\label{TIMES}}
\end{figure}

As a result we show the performance plot of the Q4 in
Fig.~\ref{TIMES}.  The total time expense (total) and the pure
communication time (ipc) for one computation of all forces are plotted
for both the standard-systolic (sys) and the hyper-systolic (hys)
algorithms.
As expected, for both methods, we find a linear increase in
communicational and a quadratic behaviour of the total time expense,
as we are working on a fixed machine size, with constant number of
processors $p$. The insertion zooms the curves in a regime of $n$ that
is common in practical computations on a machine with the number and
performance of processors as provided by the Q4 (\# of particles $<
500$ for the Q4).  We find here that both advantages of the
hyper-systolic method lead to their full pay-off: the communication
time is drastically reduced and together with the smaller amount of
computation operations, the total time expense is diminished by more
than a factor of two. For example, one interaction step on 192
particles requires 4.2 milliseconds in the systolic case compared to
1.8 milliseconds in the hyper-systolic mode.

\begin{figure}[htb]
\centerline{\epsfxsize=8cm\epsfbox{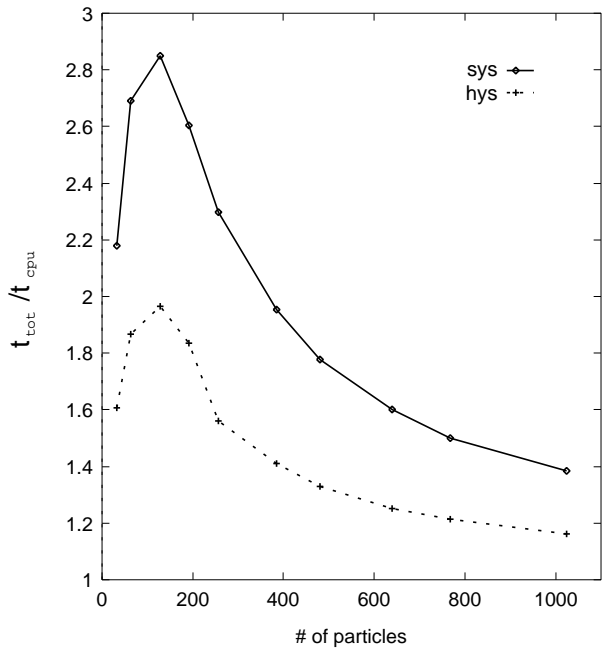}}
\caption[]{
$\frac{t_{total}}{t_{cpu}}$ as function of the number of particles on the Q4.
\label{RATIO}}
\end{figure}

Fig.~\ref{RATIO} plots the `inefficiency factor', $t_{total}/t_{cpu}$,
 for both algorithms.
The threshold behaviour of the inefficiency factor in the region
$n\leq 150$ reflects the characteristics of the Quadrics register
structure: for the registers to be filled without bubbles, one needs
at least a set of $2 \times 6$ particles, while the communication time
increases proportional to $n$.  Notice that communication plays a much
smaller r\^ole for the hyper-systolic method and even for a
granularity of $g=32$, \ie\ $n=1024$, the performance in the
standard-systolic approach is still considerably affected by
communication.

\begin{figure}[htb]
\centerline{\epsfxsize=8cm\epsfbox{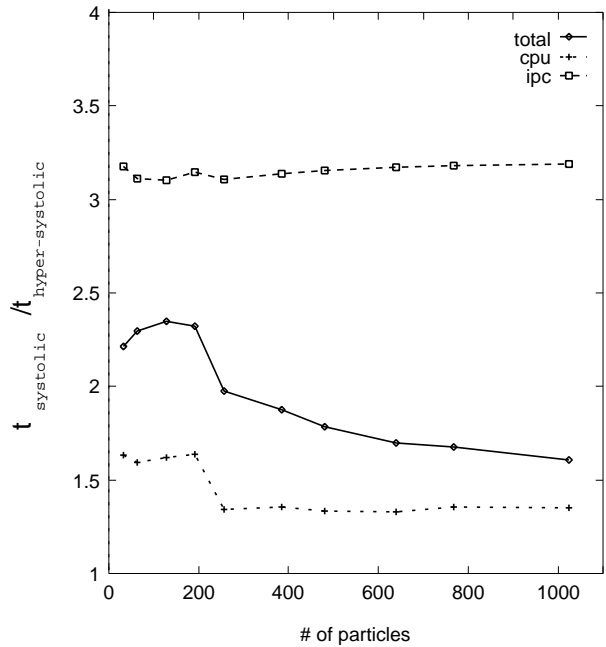}}
\caption[]{Gain factor $\frac{t_{sys}}{t_{hys}}$ for overall
 time (total), net
  computation (cpu) and interprocessor communication (ipc) times as measured
  on the Q4.
\label{TIMERATIO}}
\end{figure}

In Fig.~\ref{TIMERATIO}, we display the gain factors
$\frac{t_{sys}}{t_{hys}}$  separately, in terms of the total,
the net computational and the interprocessor communication overhead
times.
On the Q4, we actually achieve a gain factor of about $3.2$ in
communication time across the entire granularity range, to be compared
to the predicted value of $3$. The startup behaviour of the three
curves at small $n$ again reflects the filling of the registers.  The
overall gain reaches its maximum in the range of 128 to 192 particles,
whereupon the competing complexities are leading to more modest
improvements.  Yet, the hyper-systolic mode still results in an
attractive gain of $60 \%$ for 1024 particles, at granularity 32!

\begin{figure}[htb]
\centerline{\epsfxsize=8cm\epsfbox{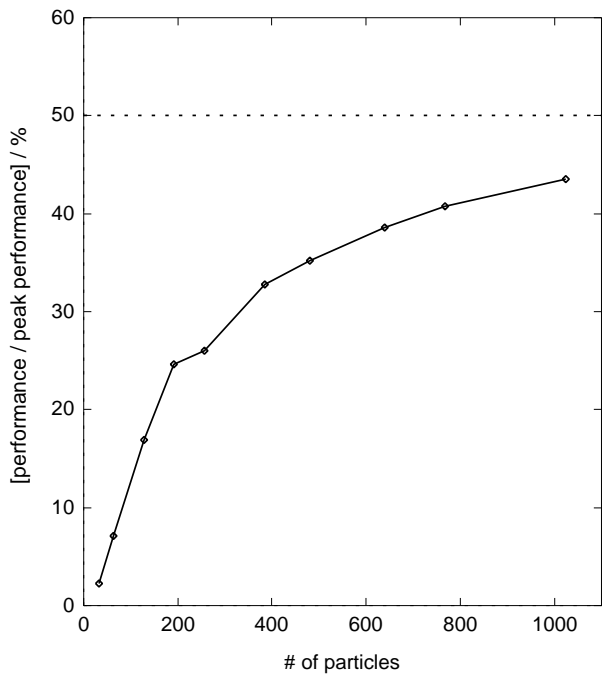}}
\caption[]{ Performance in \% of the peak performance on
  the Q4.  \label{FLOATING}} \end{figure}

Let us turn next to efficiency measurements.  Counting the computation
of an inverse square root with $18$ floating point operations, we
assign $35$ floating point operations to the computation of each
$r_{ij}$ in \eq{NEWTON} on the 3-dimensional problem.
Fig.~\ref{FLOATING} presents the flop rates as measured with the
hyper-systolic method on the Q4 (in units of its peak performance) for
the computation of the Coulomb interaction.
 Before the registers are
fully exploited ($< 150$ particles on the Q4), the performance is
small as expected.  For larger values of $n$, however, a remarkably
high performance can be achieved for the full n-body computation of
the Coulomb force in three dimensions.  In terms of the peak
performance of the Quadrics Q4, \ie\ 1.6 Gflop, nearly $50$ \% can be
reached in the limit of large $n$.  In real life, one would consider
the Q4 to be the adequate Quadrics configuration for treating 200 to
500 particles; in this range we observe efficiencies between $20$ and
$30$ \% of the peak performance.

\subsection{Performance Tests on the QH4\label{PERFORMANCEQH}}
Next we extend our performance measurements to the QH4 configuration
which is equipped with 512 processing nodes and usually configured as
$8\times 8\times8$ grid.

\begin{figure}[htb]
\centerline{\epsfxsize=8cm\epsfbox{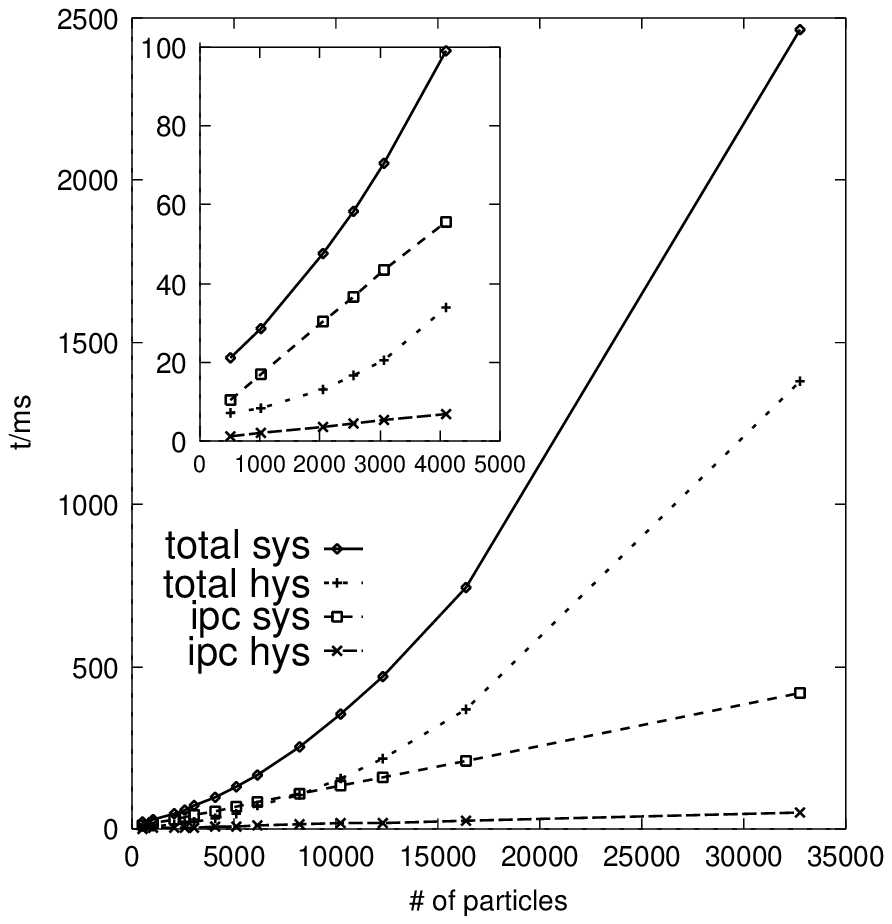}}
\caption[Times]{Total time (total) and  communication time (ipc) vs.\
  the number of particles for one interaction time-step on the
  512-nodes Quadrics QH4 for the systolic and the hyper-systolic
  method.
\label{TIMES_QH4}}
\end{figure}

As explained above , we expect an improvement in interprocessor
communication by about a factor of $8$ on the QH4 (as compared to $3$
on the 32-node Q4)\footnote{On top of that, there should be
  additional, computational gains with respect to the Q4.}.  The
results of our QH4 benchmark are displayed in Fig.~\ref{TIMES_QH4},
which is the analogue to the Q4 plot in Fig.~\ref{TIMES}.
The insertion zooms the granularity range $g\leq 8$. Characteristic
for the hyper-systolic algorithm is the fact that, even for $g=1$,
$t_{ipc}$ is one order of magnitude smaller than $t_{cpu}$. In the
whole range of granularities investigated, for the hyper-systolic
method, the size of $t_{ipc}$ is marginal, whilst the
standard-systolic computation is dominated by communication overhead,
up to the point $g=16$.

\begin{figure}[htb]
\centerline{\epsfxsize=8cm\epsfbox{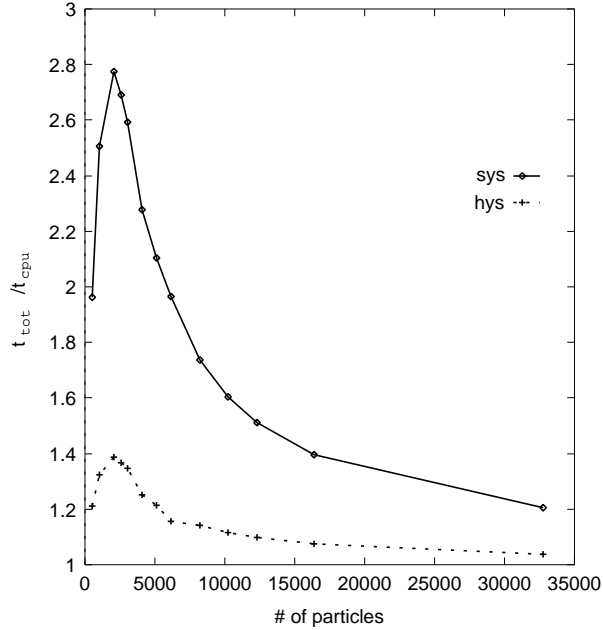}}
\caption[]{
Inefficiency factor $\frac{t_{total}}{t_{cpu}}$ as function of
the number of particles on the QH4.
\label{RATIO_QH4}}
\end{figure}

The QH4 inefficiency factor, as shown in Fig.~\ref{RATIO_QH4} is very
reminiscent to Fig.~\ref{RATIO}, featuring again the effects of
register filling in form of the characteristic threshold behaviour at
small $n$.
The important message to be drawn from this plot is the finding that
in the {\it relevant intermediate} region of say $g<16$,
hyper-systolic are by far superior to standard-systolic
implementations in performance.  Note that in the {\it asymptotic}
regime of very large granularity the machine can be viewed more and
more as an assembly of independently working processors; in this
`scalar' limit, communication ceases to dominate for both systolic
varieties, but our $n^2$-loop computation approaches the creeping
mode!

\begin{figure}[htb]
  \centerline{\epsfxsize=8cm\epsfbox{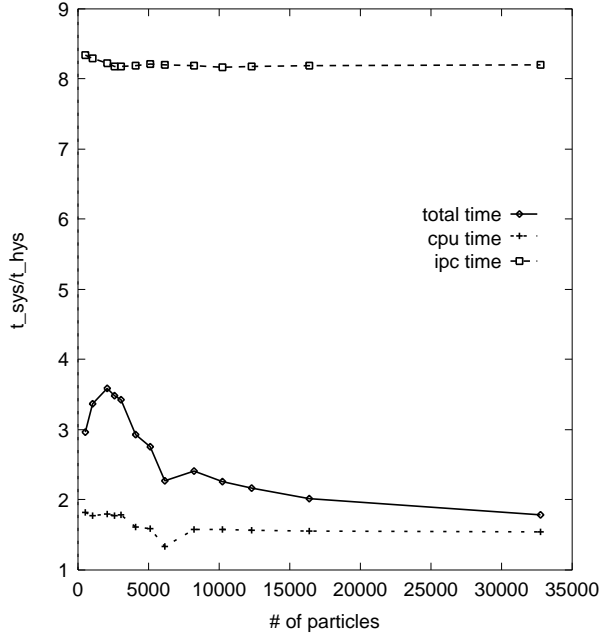}}
\caption[]{Ratio $\frac{t_{sys}}{t_{hys}}$ for total time (total), pure
computation (cpu) and interprocessor communication (ipc)
on the QH4.
\label{TIMERATIO_QH4}}
\end{figure}

In Fig.~\ref{TIMERATIO_QH4} we display the gain factor for the
hyper-systolic implementation on the QH4 for particle numbers up to
35.000.  It is not surprising after our theoretical discussion in
section \ref{IMPQH4} to find an improvement of $>8$ in the
interprocessor communication time throughout the observed $g$ range.
Here we reach a slightly improved level for the gain in CPU-time as
the computations contribute more heavily (larger $n$ as compared to
the Q4 case). The improved communication time boosts the peak in the
overall gain factor from $2.8 $ on the Q4 to $3.5$ on the QH4.

\begin{figure}[htb]
\centerline{\epsfxsize=8cm\epsfbox{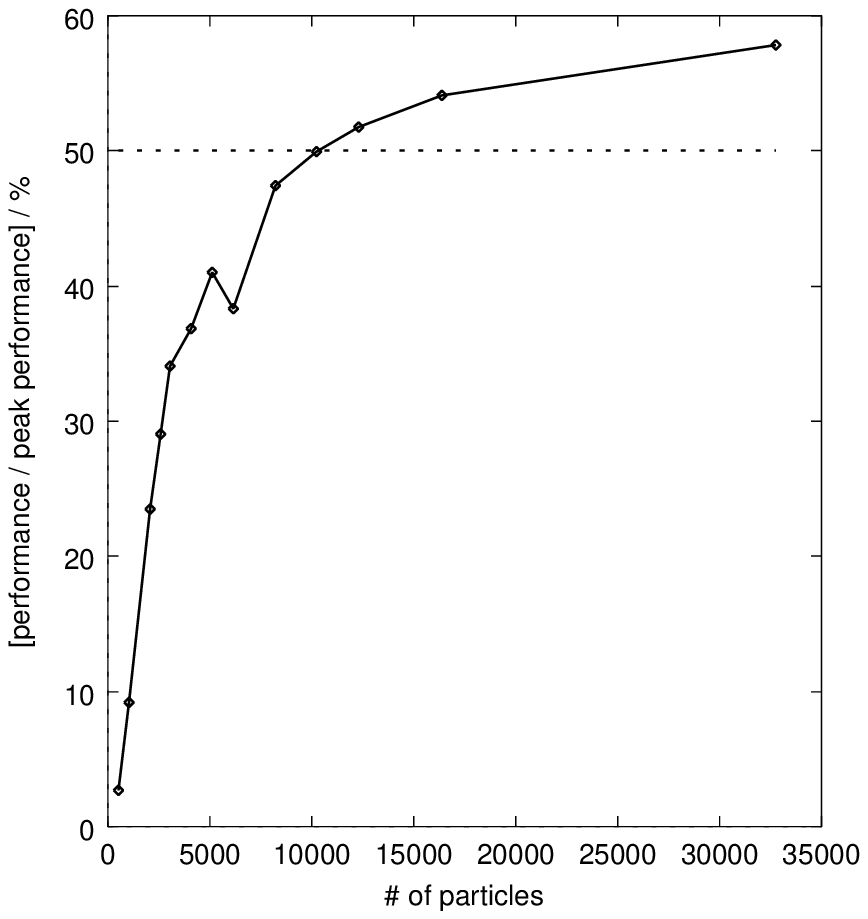}}
\caption[]{
Performance in \% of the peak performance  on the QH4.
\label{FLOATING_QH4}}
\end{figure}

Finally it is rather gratifying to find a high efficiency level for
the QH4 in terms of its peak performance, see Fig.~\ref{FLOATING_QH4}.
The efficiency of the QH4 in the hyper-systolic implementation of
molecular dynamics saturates at a value close to $60 \%$, crossing the
$50 \%$ mark at granularity $20$. Even at low granularity like
$g\approx 6$ we observe efficiencies in the ball-park $30$ to $40$ \%!
\section{Summary}

We have investigated standard-systolic and hyper-systolic computations
for a prototype $n^2$-loop problem on the massively parallel computer
Quadrics.

A next-neighbour data movement on the logical systolic chain ``GRA''
is realized in form of three interprocessor communication steps, the
hyper-systolic data movement can be performed in one interprocessor
communication step on the Q4, two steps on the $8\times 4\times 8$ QH2
and three steps on $8\times 8\times 8$ QH4, respectively.

As expected, our measurement reveals a gain factor of $\approx 3$ in
communication time compared to the standard-systolic computation, on
the Quadrics Q4.  In this way, we verified a performance of up to 750
Mflops in molecular dynamics with exact long range gravitational
forces for acceptable granularity.  Saturation would be reached at
about 50 \% of the theoretical peak performance of the machine.

Even more encouraging results were subsequently found in measurements
on the largest actually built member of the Quadrics line, the QH4,
where the cpu-power can be exploited even for small granularity.  The
performance saturates for $n > 32 k$ near $60$ \% of the peak
performance, \ie\ nearly 15.5 Gflops, for long-range Coulomb
interactions.

According to the results, on the QH4, it appears realistic to perform
molecular dynamics simulations with $O(16 k)$ particles in 6 days
compute time with a length of $10^6$ molecular dynamics time steps.

We thus conclude that the APE100/Quadrics architecture should no
longer be considered as special purpose machines for lattice QCD.
Instead they should be viewed as powerful devices that can solve just
as well many other bottleneck problems of Computational Science
belonging to the wide class of $n^2$-loop problems.

In two consecutive papers we will evaluate the impact of the
hyper-systolic approach on other bottleneck problems in HPCN such as
matrix products\cite{MATRIX} and fast Fourier transforms.

\section*{Acknowledgements}

Part of this investigation has been carried out at ENEA/Casaccia.  Th.
L., G. R. and K. S. thank Profs. N.\ Cabibbo and A.\ Mathis for their
kind hospitality. We are indebted to Dott. R. Guadagni and Dott.  R.
Canata for their support.

The work of Th.\ L.\ is supported by the Deutsche
Forschungsgemeinschaft DFG under grant No.\ Schi 257/5-1.

\end{document}